# Generalized partially bent functions

Jianqin Zhou

(Dept. of Computer Science, Anhui University of Technology, Ma'anshan 243002, P. R. China)

(E-mail: zhou63@ahut.edu.cn)

**Abstract:** Based on the definition of generalized partially bent functions, using the theory of linear transformation, the relationship among generalized partially bent functions over ring $Z_N$, generalized bent functions over ring $Z_N$ and affine functions is discussed. When N is a prime number, it is proved that a generalized partially bent function can be decomposed as the addition of a generalized bent function and an affine function. The result obtained here generalizes the main works concerning partially bent functions by Claud Carlet in [1].

Bent functions, a special class of Boolean functions, are of great interest in the fields of cryptography and communications due to their nonlinearity and stableness. However, bent functions are rare and they are neither balanced nor correlation-immune. So partially bent functions, a larger class of Boolean functions, presented by Claud Carlet in [1] to remedy the defects of bent functions. Now concepts of bent and partially bent functions have been extended onto ring $Z_N$, N is a natural number, called generalized bent functions and generalized partially bent functions over $Z_N$, respectively.

Based on the definition of generalized partially bent functions, using the theory of linear transformation, the relationship among generalized partially bent functions over ring $Z_N$, generalized bent functions over ring $Z_N$ and affine functions is discussed. When N is a prime number, such as N=2, it is proved that a generalized partially bent function can be decomposed as the addition of a generalized bent function and an affine function. The result obtained here generalizes the main works concerning partially bent functions by Claud Carlet in [1]. With these new results, we can easily understand and construct partially bent functions and generalized partially bent functions.

## 1. Preliminaries

Let $Z_N$, where N>1 is an integer, be a residue ring. If N is a prime number, then $Z_N$ is a Galois field, denoted by $F_N$. Let $Z_N^n$ be the set of all vectors with n coordinates, where each coordinate takes a value from $Z_N$. If N is a prime number, then $Z_N^n$ is a linear space of dimension n over $F_N$, denoted by $F_N^n$. Let f: $Z_N^n \to Z_N$ be a multivalued logical function.

Let $a = (a_1, \ldots, a_n) \in Z_N^n$ and $x = (x_1, \ldots, x_n) \in Z_N^n$, the inner product of $a$ and x is defined as $a \cdot x = a_1 x_1 \oplus \ldots \oplus a_n x_n$.

**Definition 1.1.** The Chrestenson cyclic spectrum is defined as follows:

$S_{(f)}(w) = N^{-n} \sum_{x \in Z_N^n} u^{f(x)} u^{-w \bullet x}$, where $w \in Z_N^n$, $u = \exp(2\pi\sqrt{-1}/N)$;

The self-correlation function is defined as follows:

$C_f(s) = \sum_{x \in Z_N^n} u^{f(x+s)-f(x)}$, where $s \in Z_N^n$, $u = \exp(2\pi\sqrt{-1}/N)$.

We will denote by |X| the module of X, where X is real number or complex number.

**Definition 1.2.** The function f(x) is called generalized bent if

$|S_{(f)}(w)|^2 = N^{-n}$ for all $w \in Z_N^n$;

The function f(x) is called generalized partially bent if

$(N^n - N_{Cf})(N^n - N_{S(f)}) = N^n$, where $N_{Cf} = |\{s \in Z_N^n | C_f(s) = 0\}|$ and $N_{S(f)} = |\{s \in Z_N^n | S_{(f)}(s) = 0\}|$.

The following facts are well known.

**Lemma 1.1**. For any function f(x): $Z_N^n \to Z_N$, $\sum_{x \in Z_N^n} |S_{(f)}(x)|^2 = 1$.

**Lemma 1.2**. For any function f(x): $Z_N^n \to Z_N$, $N^{-n} \sum_{x \in Z_N^n} C_f(x) u^{-w \bullet x} = N^n |S_{(f)}(w)|^2$,

There are similar definitions and lemmas concerning bent functions and partially bent functions.

## 2. Main theorems

The following theorem 2.1 is needed by other theorems.

**Theorem 2.1**. Let f(x): $Z_N^n \to Z_N$ be a multivalued logical function and A be an inverse matrix over $Z_N$ and g(x)=f(xA),

then $C_g(v) = C_f(vA)$; $S_{(g)}(w) = S_{(f)}(w(A^{-1})^t)$.

**Proof:** Let $e_i (1 \leq i \leq n)$ denote a vector of $Z_N^n$ such that the ith coordinate is 1 while all other coordinates are 0.

Let $(\alpha_1, \alpha_2, \cdots, \alpha_n)^t = A \begin{pmatrix} e_1 \\ e_2 \\ . \\ . \\ . \\ e_n \end{pmatrix}$, where $a^t$ denotes the transpose of vector $\alpha$.

Since A is an inverse matrix, $e_i (1 \leq i \leq n)$ can be expressed as a linear combination of $(\alpha_1, \alpha_2, \cdots, \alpha_n)$, thus $(\alpha_1, \alpha_2, \cdots, \alpha_n)$ is a radix of $Z_N^n$.

Let $y = (y_1, y_2, \cdots, y_n)$ denote $y_1 \alpha_1 + y_2 \alpha_2 + \cdots + y_n \alpha_n$ of $Z_N^n$.

Note that $y_1 \alpha_1 + y_2 \alpha_2 + \cdots + y_n \alpha_n = (y_1, y_2, \cdots, y_n)(\alpha_1, \alpha_2, \cdots, \alpha_n)^t = (y_1, y_2, \cdots, y_n) A \begin{pmatrix} e_1 \\ e_2 \\ . \\ . \\ . \\ e_n \end{pmatrix}$,

Thus $f(y_1 \alpha_1 + y_2 \alpha_2 + \cdots + y_n \alpha_n) = f(yA) = g(y_1, y_2, \cdots, y_n)$,

$C_g(v) = \sum_{y \in Z_N^n} u^{g(y+v)-g(y)} = \sum_{y \in Z_N^n} u^{f(yA+vA)-f(yA)} = \sum_{w \in Z_N^n} u^{f(w+vA)-f(w)} = C_f(vA)$,

In fact, $v = (v_1, v_2, \cdots, v_n)$ denotes $v_1 \alpha_1 + v_2 \alpha_2 + \cdots + v_n \alpha_n = (v_1, v_2, \cdots, v_n) A \begin{pmatrix} e_1 \\ e_2 \\ . \\ . \\ . \\ e_n \end{pmatrix}$.

$S_{(g)}(w) = N^{-n} \sum_{x \in Z_N^n} u^{g(x)} u^{-xw^t} = N^{-n} \sum_{x \in Z_N^n} u^{f(xA)} u^{-xw^t} = N^{-n} \sum_{y \in Z_N^n} u^{f(y)} u^{-yA^{-1}w^t} = S_{(f)}(w(A^{-1})^t)$.

This completes the proof.∎

We now discuss the generalized partially bent functions.

**Theorem 2.2.** Let $f(x): Z_N^n \rightarrow Z_N$, $N_{Cf} = |\{s \in Z_N^n | C_f(s) = 0\}|$, $N_{S(f)} = |\{s \in Z_N^n | S_{(f)}(s) = 0\}|$, then

(I)    $(N^n - N_{Cf})(N^n - N_{S(f)}) \geq N^n$;

(II)   $f(x)$ is generalized partially bent, namely $(N^n - N_{Cf})(N^n - N_{S(f)}) = N^n$, if and only if the following conditions are true:

There exists $t \in Z_N^n$, such that for any $s \in Z_N^n$, $C_f(s) = 0$ or $C_f(s) = u^{-s \cdot t} N^n$, and $|S_{(f)}(w)|^2$ is a constant when $w \in Z_N^n$ and $S_{(f)}(w) \neq 0$.

**Proof:**

(a). Since $|C_f(s)| \leq N^n$, hence,

$N^n - N_{Cf} = |\{s \in Z_N^n | C_f(s) \neq 0\}| \geq \sum_{s \in Z_N^n} |C_f(s)| / N^n \geq |\sum_{s \in Z_N^n} C_f(s)| / N^n$

$= |\sum_{s \in Z_N^n} \sum_{x \in Z_N^n} u^{f(x+s)-f(x)}| / N^n = |\sum_{x \in Z_N^n} u^{-f(x)} \sum_{s \in Z_N^n} u^{f(x+s)}| / N^n$

$$=|S_{(f)}(0)| \cdot |\sum_{x \in Z_N^n} u^{-f(x)}| = |S_{(f)}(0)| \cdot |\sum_{x \in Z_N^n} u^{f(x)}| = N^n |S_{(f)}(0)|^2.$$

Let $f_1(x) = f(x) + t \cdot x$, then,

$$C_{f_1}(s) = \sum_{x \in Z_N^n} u^{f(x+s)-f(x)} u^{s \cdot t} = u^{s \cdot t} C_f(s),$$

$$S_{(f_1)}(s) = N^{-n} \sum_{x \in Z_N^n} u^{f(x)} u^{t \cdot x - s \cdot x} = S_{(f)}(s-t).$$

Thus, for any $t \in Z_N^n$, $N^n - N_{Cf} = N^n - N_{Cf_1} \geq N^n |S_{(f_1)}(0)|^2 = N^n |S_{(f)}(-t)|^2 = N^n |S_{(f)}(t)|^2$

Therefore, $N^n - N_{Cf} \geq N^n \max\{|S_{(f)}(t)|^2 | t \in Z_N^n\}$

Since $N^n - N_{S(f)} \geq \sum_{t \in Z_N^n} |S_{(f)}(t)|^2 / \max\{|S_{(f)}(t)|^2 | t \in Z_N^n\} = 1/\max\{|S_{(f)}(t)|^2 | t \in Z_N^n\}$

Then we have $(N^n - N_{Cf})(N^n - N_{S(f)}) \geq N^n$. We have completed the proof of part one.

(b). If $(N^n - N_{Cf})(N^n - N_{S(f)}) = N^n$, then

$N^n - N_{Cf} = N^n \max\{|S_{(f)}(t)|^2 | t \in Z_N^n\}$ and $N^n - N_{S(f)} = 1/\max\{|S_{(f)}(t)|^2 | t \in Z_N^n\}$

Suppose $t \in Z_N^n$, and $|S_{(f)}(t)|^2 = \max\{|S_{(f)}(u)|^2 | u \in Z_N^n\}$, let $f_1(x) = f(x) + t \cdot x$, then

$$\sum_{s \in Z_N^n} C_{f_1}(s) = N^{2n} |S_{(f_1)}(0)|^2 = N^{2n} |S_{(f)}(t)|^2 = N^n (N^n - N_{Cf}) = N^n (N^n - N_{Cf_1}) = \sum_{s: C_{f1}(s) \neq 0} N^n$$

Since $|C_{f_1}(s)| \leq N^n$, hence $C_{f_1}(s) = 0$ or $N^n$, namely $C_f(s) = 0$ or $C_f(s) = u^{-s \cdot t} N^n$;

Consider $N^n - N_{S(f)} = |\{s \in Z_N^n | S_{(f)}(s) \neq 0\}| = \sum_{t \in Z_N^n} |S_{(f)}(t)|^2 / \max\{|S_{(f)}(t)|^2 | t \in F_N^n\}$, we know that $|S_{(f)}(w)|^2$ is a constant when $w \in Z_N^n$ and $S_{(f)}(w) \neq 0$.

This completes the necessity proof of part two.

(c). Suppose that there exists $t \in Z_N^n$, such that for any $s \in Z_N^n$, $C_f(s) = 0$ or $C_f(s) = u^{-s \cdot t} N^n$, and $|S_{(f)}(w)|^2$ is a constant when $w \in Z_N^n$ and $S_{(f)}(w) \neq 0$.

Let $E = \{s \in Z_N^n | C_f(s) = u^{-s \cdot t} N^n\}$, $f_1(x) = f(x) + t \cdot x$, then $C_{f_1}(s) = u^{s \cdot t} C_f(s) = 0$ or $N^n$, hence $E = \{s \in Z_N^n | C_{f_1}(s) = N^n\}$.

Let $E^\perp = \{x \in Z_N^n | \text{for any } y \in E, y \cdot x = 0\}$, then for any $v \in E^\perp$, by Lemma 1.2, we have

$$|S_{(f_1)}(v)|^2 = \frac{1}{N^{2n}} \sum_{w \in Z_N^n} C_{f_1}(w) u^{-w \cdot v} = \frac{1}{N^n} \sum_{w \in E} u^{-w \cdot v} = \frac{1}{N^n} |E|.$$

As $f_1(x) = f(x) + t \cdot x$, so $S_{(f_1)}(s) = S_{(f)}(s-t)$; consider $|S_{(f)}(w)|^2$ is a constant when $w \in Z_N^n$ and $S_{(f)}(w) \neq 0$, hence $|S_{(f_1)}(w)|^2$ is a constant when $w \in Z_N^n$ and $S_{(f_1)}(w) \neq 0$.

When $v \notin E^\perp$, the real part of $\sum_{w \in E} u^{-w \cdot v} < |E|$,

Therefore $|S_{(f_1)}(v)|^2 = \frac{1}{N^n} \sum_{w \in E} u^{-w \cdot v} \neq \frac{1}{N^n} |E|$, so $|S_{(f_1)}(v)|^2 = 0$.

By Lemma 1.1, $\sum_{w \in Z_N^n} |S_{(f_1)}(w)|^2 = 1$, so $|E^\perp| \cdot \frac{1}{N^n} |E| = 1$, we have $|E^\perp| = \frac{N^n}{|E|}$.

Thus $(N^n - N_{Cf}) = |E|$, $(N^n - N_{S(f)}) = \frac{N^n}{|E|}$, namely $(N^n - N_{Cf})(N^n - N_{S(f)}) = N^n$, $f(x)$ is generalized partially bent.

This completes the sufficiency proof of part two. ∎

**Theorem 2.3**. Let $f(w): Z_N^n \to Z_N$ be generalized partially bent, then there exist a subgroup $E$ in the additive group $Z_N^n$, such that, there exists $t \in Z_N^n$ and for any $x \in E$, $y \in Z_N^n \backslash E = \{x | x \in Z_N^n \text{ but } x \notin E\}$, satisfying $f(x+y) = f(y) - t \cdot x$.

**Proof:** Since $f(x)$ is generalized partially bent, let $E = \{s \in Z_N^n | C_f(s) = u^{-s \cdot t} N^n\}$, here $t \in Z_N^n$ and the definition of $t$ is from theorem 2.2; let $f_1(x) = f(x) + t \cdot x$, then $C_{f_1}(s) = u^{s \cdot t} C_f(s)$, $E = \{s \in Z_N^n | C_{f_1}(s) = N^n\}$.

Since $C_{f_1}(s) = \sum_{x \in Z_N^n} u^{f_1(x+s)-f_1(x)}$, hence α ∈ E if and only if that for any x∈ $Z_N^n$, $f_1(x+α) - f_1(x) = 0$, namely f(x+α)+t•(x+α) - f(x) - t•x = 0, so f(x+α)+t•α = f(x),

Suppose α, β ∈ E, then for any x ∈ $Z_N^n$, f(x+α+β)+t•(α+β) = f(x+α+β)+t•β + t•α = f(x+α) + t•α = f(x), thus α + β ∈ E,

For any k ∈ $Z_N$, it is easy to show that kα ∈ E. Since (N-1)α ∈ E and α +(N-1)α =0∈ E, so the inverse element of α still belongs to E.

Therefore, E is a subgroup of the additive group $Z_N^n$. It is obvious that for any x ∈ E, y ∈ $Z_N^n$ \E, satisfying f(x+y)=f(y)-t•x. This completes the proof.■

**Lemma 2.1.** Let $m_i$ = min{ $α_i > 0$ | ($α_1$, ···, $α_i$, ···, $α_n$) ∈ E}, 1 ≤ i ≤ n, the definition of E is from theorem 2.3, if $m_i$ is not defined then let $m_i$ = 0. Our conclusion is that if $m_i$ is neither 0 nor 1, then $m_i$ must be a factor of N but $m_i$ is neither N nor 1.

**Proof:** Suppose that $m_i$ is neither 0 nor 1, let q=($m_i$, N), then q>0 and there exists an integer r and an integer s, such that

q = r$m_i$ + sN, namely q = r$m_i$ (mod N),

From the definition of $m_i$, we have q = $m_i$, thus $m_i$ is a factor of N but $m_i$ is neither N nor 1.

This completes the proof.■

For the convenience of discussions, we give a new definition.

**Definition 2.1.** Let f(w): $Z_N^n \to Z_N$ be generalized partially bent, E={s∈ $Z_N^n$ | $C_f(s) = u^{-s \cdot t} N^n$ }, t∈ $Z_N^n$, $m_i$ = min{ $α_i > 0$ |($α_1$,···, $α_i$,···, $α_n$) ∈ E}, 1 ≤ i ≤ n, if $m_i$ is not defined then let $m_i$ = 0. If $m_i \ne 1$ for any i, 1 ≤ i ≤ n, then we call f(w) as pure generalized partially bent.

If N is a prime number, as N has only factor 1 and N, then if f(w) is pure generalized partially bent, namely $m_i$ =0 for any i, 1 ≤ i ≤ n, we have E={0}, thus f(w) must be generalized bent.

The following theorem 2.4 is our main result.

**Theorem 2.4.** f(x): $Z_N^n \to Z_N$ be generalized partially bent, but not pure generalized partially bent, if and only if f(x) is equivalent to the addition of pure generalized partially bent g(y): $Z_N^{n-m} \to Z_N$ and affine function -$t_1$ • α : $Z_N^m \to Z_N$, here m is a positive integer, and f(x) is equivalent to h(x) means that there exists an inverse matrix A over $Z_N$, such that h(x)=f(xA).

**Proof:** Let f(w): $Z_N^n \to Z_N$ be generalized partially bent, $E_f$ ={s∈ $Z_N^n$ | $C_f(s) = u^{-s \cdot t} N^n$ }, t∈ $Z_N^n$ and the definition of t is from theorem 2.2. If f(x) is not pure generalized partially bent, then there exists $m_i$ =1 and the definition of $m_i$ is from theorem 2.3. That is to say, there exist ($α_1$,···, $α_i$,···, $α_n$) ∈ $E_f$, $α_i$ =1.

Let $A_1 = \begin{pmatrix} 1 & 0 & \mathbf{L} & & 0 \\ 0 & 1 & \mathbf{L} & & 0 \\ \mathbf{L} & & \mathbf{L} & & \mathbf{L} \\ a_1 & a_2 & \mathbf{L} & a_i \mathbf{L} & a_n \\ \mathbf{L} & & \mathbf{L} & & \mathbf{L} \\ 0 & 0 & \mathbf{L} & & 1 \end{pmatrix}$, then $A_1^{-1} = \begin{pmatrix} 1 & 0 & \mathbf{L} & & 0 \\ 0 & 1 & \mathbf{L} & & 0 \\ \mathbf{L} & & \mathbf{L} & & \mathbf{L} \\ -a_1 & -a_2 & \mathbf{L} & a_i \mathbf{L} & -a_n \\ \mathbf{L} & & \mathbf{L} & & \mathbf{L} \\ 0 & 0 & \mathbf{L} & & 1 \end{pmatrix}$, where $α_i$ =1.

$A_2 = \begin{pmatrix} 0 & 0 & \mathbf{L} & 1 \mathbf{L} & 0 \\ 1 & 0 & \mathbf{L} & & 0 \\ \mathbf{L} & & \mathbf{L} & & \mathbf{L} \\ 0 & 0 & \mathbf{L} 1 & 0 \mathbf{L} & 0 \\ \mathbf{L} & & \mathbf{L} & & \mathbf{L} \\ 0 & 0 & \mathbf{L} & & 1 \end{pmatrix}$,

where $A_2$ is obtained by move the ith row of a unit matrix to the top row, then $A_2^{-1} = A_2^t$.

Let $A = A_2 A_1$, $(\beta_1, \beta_2, \cdots, \beta_n)^t = A \begin{pmatrix} e_1 \\ e_2 \\ \cdot \\ \cdot \\ \cdot \\ e_n \end{pmatrix}$,

where $e_i$ ($1 \leq i \leq n$) denotes a vector of $Z_N^n$ such that the ith coordinate is 1 while all other coordinates are 0.

Since A is an inverse matrix, $e_i$ ($1 \leq i \leq n$) can be expressed as a linear combination of $(\beta_1, \beta_2, \cdots, \beta_n)$, thus $(\beta_1, \beta_2, \cdots, \beta_n)$ is a radix of $Z_N^n$, where $\beta_1 = \alpha_1 e_1 + \cdots + \alpha_i e_i + \cdots + \alpha_n e_n$, $\beta_2 = e_1, \cdots, \beta_i = e_{i-1}, \beta_{i+1} = e_{i+1}, \cdots, \beta_n = e_n$.

Let $g(y) = f(yA)$, by theorem 2.1, we know that $|S_{(f)}(w)|^2$ is a constant when $w \in Z_N^n$ and $S_{(f)}(w) \neq 0$; $C_g(s) = 0$ or $C_g(s) = C_f(sA) = u^{-sAt^t} N^n = u^{-s(tA^t)^t} N^n$. By theorem 2.2, $g(y)$ is till generalized partially bent.

Let $t' = tA^t$, $g_1(x) = g(x) + t' \cdot x$, then $C_{g_1}(s) = u^{s \cdot t'} C_g(s) = u^{s \cdot At^t} C_g(s) = N^n$, here $s \bullet t = s \bullet t^t$.

Let $E_{g1} = \{ s \in Z_N^n \mid C_{g_1}(s) = N^n \}$ and $Z_N^1$ denote the generated subgroup by $\beta_1$, since $(\alpha_1, \cdots, \alpha_i, \cdots, \alpha_n) \in E_f$ and $\beta_1 = \alpha_1 e_1 + \cdots + \alpha_i e_i + \cdots + \alpha_n e_n$, hence $\beta_1 \subset E_{g1}$, it follows that $Z_N^1 \subset E_{g1}$.

Let $(Z_N^1)^\perp = \{ x \in Z_N^n \mid \text{for any } y \in Z_N^1, y \bullet x = 0\}$, then $(Z_N^1)^\perp = Z_N^{n-1}$, it is obvious that $Z_N^n = Z_N^1 \oplus Z_N^{n-1}$, where $\oplus$ denotes the inner direct product of two subgroups.

For any $\alpha \in Z_N^1$ and $y \in Z_N^{n-1}$, $g_1(y + \alpha) - g_1(y) = 0$, namely $g(y + \alpha) + t' \bullet (y + \alpha) - g(y) - t' \bullet x = 0$, thus $g(y + \alpha) + t' \bullet \alpha = g(y)$.

We know that there exist $t_1 \in Z_N^1$ and $t_2 \in Z_N^{n-1}$, such that $t' = t_1 + t_2$, therefore, for any $\alpha \in Z_N^1$ and $y \in Z_N^{n-1}$, $g(y) = g(y + \alpha) + t' \bullet \alpha = g(y + \alpha) + (t_1 + t_2) \bullet \alpha = g(y + \alpha) + t_1 \bullet \alpha$.

We have $g(y + \alpha) = g(y) - t_1 \bullet \alpha$, and $g(y + \alpha)$ is an affine function restricted within $Z_N^1$.

Furthermore, we now prove $g(y + \alpha)$ is generalized partially bent restricted within $Z_N^{n-1}$.

Take $v \in Z_N^{n-1}$, for any $w \in Z_N^n$, there exit $x \in Z_N^1$ and $y \in Z_N^{n-1}$, such that $w = x + y$, hence $g(w + v) - g(w) = g(x + y + v) - g(x + y) = g(y + v) - t_1 \bullet x - g(y) + t_1 \bullet x = g(y + v) - g(y)$, thus,

$$C_g(v) = \sum_{w \in Z_N^n} u^{g(w+v) - g(w)} = \sum_{x \in Z_N^1} \sum_{y \in Z_N^{n-1}} u^{g(y+v) - g(y)} = N \sum_{y \in Z_N^{n-1}} u^{g(y+v) - g(y)},$$

$$\therefore \sum_{y \in Z_N^{n-1}} u^{g(y+v) - g(y)} = 0 \text{ or } \sum_{y \in Z_N^{n-1}} u^{g(y+v) - g(y)} = C_g(v)/N = u^{-v \cdot t'} N^{n-1} = u^{-v \cdot t_2} N^{n-1}, t_2 \in Z_N^{n-1};$$

We know that, for any $v, w \in Z_N^n$, there exit $v_1, x \in Z_N^1$ and $v_2, y \in Z_N^{n-1}$, such that $v = v_1 + v_2$, $w = x + y$, $g(w) - w \bullet v = g(y) - t_1 \bullet x - (x + y) \bullet (v_1 + v_2) = g(y) - y \bullet v_2 - (t_1 + v_1) \bullet x$, thus,

$$S_{(g)}(v) = N^{-n} \sum_{w \in Z_N^n} u^{g(w) - w \cdot v} = \sum_{x \in Z_N^1} \sum_{y \in Z_N^{n-1}} u^{g(y) - y \cdot v_2 - (t_1 + v_1) \cdot x} = N^{-1} \sum_{x \in Z_N^1} u^{-(t_1 + v_1) \cdot x} \; N^{-n+1} \sum_{y \in Z_N^{n-1}} u^{g(y) - y \cdot v_2}$$

Let $v_1 = -t_1$, then $S_{(g)}(v) = N^{-n+1} \sum_{y \in Z_N^{n-1}} u^{g(y) - y \cdot v_2} = S_{(g)}(v_2)$ restricted within $Z_N^{n-1}$.

Since $|S_{(g)}(v)|^2$ is a constant when $v \in Z_N^n$ and $S_{(g)}(v) \neq 0$,

$\therefore |S_{(g)}(v_2)|^2$ restricted within $Z_N^{n-1}$ is a constant when $v_2 \in Z_N^{n-1}$ and $S_{(g)}(v_2) \neq 0$

From theorem 2.2, we know that $g(y)$ is generalized partially bent restricted within $Z_N^{n-1}$.

Therefore, the equality $g(y + \alpha) = g(y) - t_1 \bullet \alpha$ means that g is the addition of the generalized partially bent function restricted within $Z_N^{n-1}$ and the affine function restricted within $Z_N^1$.

If g restricted within $Z_N^{n-1}$ is not pure generalized partially bent, then repeat the process above we can obtain another unitary affine function.

Therefore, we can conclude that after m decompositions, where m is a positive integer, f(x) is equivalent to the addition of the generalized partially bent function g(y): $Z_N^{n-m} \to Z_N$ and the affine function $-t_1 \bullet \alpha : Z_N^m \to Z_N$.

The sufficiency of the theorem is obvious. This completes the proof.∎

If N is a prime number, then $F_N$ is a Galois field, and $F_N^n$ is a linear space of dimension n over $F_N$. Since N has only factor 1 and N, from theorem 2.4, we have the following theorem.

**Theorem 2.5**. Let N be a prime number, and f(x): $F_N^n \to F_N$ be generalized partially bent, but not generalized bent, if and only if f(x) is equivalent to the addition of generalized bent g(y): $F_N^{n-m} \to F_N$ and affine function $-t_1 \bullet \alpha : F_N^m \to F_N$, here m is a positive integer, and f(x) is equivalent to h(x) means that there exists an inverse matrix A over $F_N$, such that h(x)=f(xA).

The following is an example of pure generalized partially bent functions.

**Example 2.1.** Let $I_{\{1,3\}}(x) = \begin{cases} 1, x=1,3 \\ 0, x=0,2 \end{cases}$, and f(x,y)= $I_{\{1,3\}}(x) + I_{\{1,3\}}(y) + x + y$.

We first show that f(x,y) is a generalized partially bent function.

Here N=4, u= $\exp(2p\sqrt{-1}/N)$= $\exp(p\sqrt{-1}/2)$=$\cos p/2 + i\sin p/2$=i.

Obviously, $C_f(0,0) = \sum_{(x,y)\in Z_4^2} u^{f(x+0,y+0)-f(x,y)} = 16$.

$C_f(0,2) = \sum_{(x,y)\in Z_4^2} u^{f(x+0,y+2)-f(x,y)}$

=$i^{f(0,2)-f(0,0)} + i^{f(0,3)-f(0,1)} + i^{f(0,0)-f(0,2)} + i^{f(0,1)-f(0,3)}$
+ $i^{f(1,2)-f(1,0)} + i^{f(1,3)-f(1,1)} + i^{f(1,0)-f(1,2)} + i^{f(1,1)-f(1,3)}$
+ $i^{f(2,2)-f(2,0)} + i^{f(2,3)-f(2,1)} + i^{f(2,0)-f(2,2)} + i^{f(2,1)-f(2,3)}$
+ $i^{f(3,2)-f(3,0)} + i^{f(3,3)-f(3,1)} + i^{f(3,0)-f(3,2)} + i^{f(3,1)-f(3,3)}$
=$i^{2-0} + i^{0-2} + i^{0-2} + i^{2-0}$
+$i^{0-2} + i^{2-0} + i^{2-0} + i^{0-2}$
+$i^{0-2} + i^{2-0} + i^{2-0} + i^{0-2}$
+$i^{2-0} + i^{0-2} + i^{0-2} + i^{2-0}$=-16,

Similarly, $C_f(2,0)$=-16, $C_f(2,2)$=16.

Thus $C_f(s) = u^{-s\cdot(1,1)} 4^2$, $s \in \{0,2\} \times \{0,2\}$. It is easy to verify that $C_f(s)$=0 for $s \notin \{0,2\} \times \{0,2\}$.

$S_{(f)}(1,1) = \frac{1}{16} \sum_{(x,y)\in Z_4^2} u^{f(x,y)} u^{-x-y}$

$\frac{1}{16}(i^{f(0,0)-0} + i^{f(0,1)-1} + i^{f(0,2)-2} + i^{f(0,3)-3}$
+ $i^{f(1,0)-1} + i^{f(1,1)-2} + i^{f(1,2)-3} + i^{f(1,3)-0}$
+ $i^{f(2,0)-2} + i^{f(2,1)-3} + i^{f(2,2)-0} + i^{f(2,3)-1}$
+ $i^{f(3,0)-3} + i^{f(3,1)-0} + i^{f(3,2)-1} + i^{f(3,3)-2})$
=$\frac{1}{16}(i^{0-0} + i^{2-1} + i^{2-2} + i^{0-3}$
+ $i^{2-1} + i^{0-2} + i^{0-3} + i^{2-0}$
+ $i^{2-2} + i^{0-3} + i^{0-0} + i^{2-1}$
+ $i^{0-3} + i^{2-0} + i^{2-1} + i^{0-2}) = \frac{1}{2}i$

Similarly, $S_{(f)}(1,3) = S_{(f)}(3,1) = S_{(f)}(3,3) = \frac{1}{2}i$.

Since $\sum_{(x,y)\in Z_4^2} |S_{(f)}(x,y)|^2 = 1$, thus $S_{(f)}(v)$=0 for $v \notin \{1,3\} \times \{1,3\}$.

By definition 1.2, f(x,y) is a generalized partially bent function.

Due to E={$s \in Z_4^2 | C_f(s) = u^{-s\cdot t} 4^2$}={0,2}×{0,2}, and $m_1 = m_2 = 2$, thus f(x,y) is a pure generalized partially bent function.

Obviously, $Z_4^2$=({0,2}×{0,2}) ⊕ ({0,1}×{0,1}), for example, (2,3)=(2,2)+(0,1); (3,3)=(2,2)+(1,1).

By theorem 2.3, for any $w \in Z_4^2$, there exist $w_1 \in \{0,2\} \times \{0,2\}$ and $w_2 \in \{0,1\} \times \{0,1\}$, such that $w = w_1 + w_2$ and $f(w_1 + w_2) = f(w_2) - (1,1) \cdot w_1$.

## 3. Remarks

Since 2 is a prime number, $F_2$ is a Galois field, thus the results obtained here can be applied to partially bent functions, we have that a partially bent function can be decomposed as the addition of a bent function and an affine function. So the result has generalized the main works concerning partially bent functions by Claud Carlet.

Based on the definition of generalized partially bent functions, using the theory of linear transformation, the relationship between generalized partially bent functions over ring $Z_N$ and generalized bent functions over ring $Z_N$ is discussed. As there have been many results about bent functions and generalized bent functions, and the structure of affine functions is very simple, so it is easy to analyse or construct partially bent functions and generalized partially bent functions with the results obtained here.

A method to decompose the non-pure generalized partially bent function is introduced. However, how to decompose the pure generalized partially bent function, or how to analyse the structure of pure generalized partially bent function, is still an open problem.